\documentclass[pra,aps, english, twocolumn]{revtex4}

\usepackage{graphicx}
\usepackage{amssymb, amsmath, color, rotating}
\usepackage{bm}

\begin{document}

\title{Bosons with long range interactions on two-leg ladders in artificial magnetic fields}

\author{Stefan S. Natu}

\email{snatu@umd.edu}

\affiliation{Condensed Matter Theory Center and Joint Quantum Institute, Department of Physics, University of Maryland, College Park, Maryland 20742-4111 USA}

\begin{abstract}
Motivated by experiments exploring the physics of neutral atoms in artificial magnetic fields, we study the ground state of bosons interacting with long range \textit{dipolar} interactions on a two-leg ladder. We focus on the limit where the number of particles per site is large, and the interactions are weak. Using two complimentary variational approaches, we find rich physics driven by the long range forces.  Generically, long range interactions tend to destroy the Meissner phase in favor of modulated density wave phases. For example, nearest neighbor interactions produce an \textit{interleg charge density wave} phase, where the total density remains uniform, but the density on each leg of the ladder is modulating in space, out-of-phase with one another. \textit{Next} nearest neighbor interactions lead to a fully modulated biased ladder phase, where all the particles are on one leg of the ladder. This state simultaneously breaks $Z_{2}$ reflection symmetry and $U(1)$ translation symmetry. For values of the flux near $\phi = \pi$, we find a switching effect for arbitrarily weak interactions, where the density is modulated, but the chiral current changes sign on every plaquette. Arbitrarily weak \textit{attractive} interactions along the rungs destroy the Meissner phase completely, in favor of a modulated density wave phase. Remarkably, varying the rung to ladder hopping produces a cascade of first order transitions between modulated density wave states with different wave-vectors, which manifests itself as discrete jumps in the chiral current. Polarizing the dipoles along the ladder direction yields a region of phase space where a stable biased ladder phase occurs even at arbitrarily weak rung hopping. We discuss experimental consequences of our work, and draw connections between our work and recent experiments on cold atoms in synthetic dimensions. 

\end{abstract}
\maketitle

\section{Introduction}

The study of interacting bosons under rotation is a fundamental and rich problem: for weak rotation and weak interactions, one finds a rich array of vortex phases, whereas at strong rotation and strong interactions, one obtains bosonic analogues of the fractional quantum Hall effect \cite{Cooper2008}. More recently, it has been realized that synthetic magnetic fields or spin-orbit coupling generated by coupling atoms to Raman lasers, can mimic rotation, in that it can lead to ``kinetic frustration" by flattening the single particle band or introducing multiple degenerate minima in the band structure \cite{Spielman2011, Atala2014, OurNature, Ketterle2013, Bloch2013, Dalibard2011, Stuhl15, Fallani15}. The presence of a large single-particle degeneracy implies that interaction effects are crucial in determining the ground state (see Ref.~\cite{Zhai2012}, and references therein). The recent cooling of polar molecules \cite{Ni2008, Aikawa2010, Deiglmayr2008}, magnetic \cite{Lu2012, Pfau2007, Aikawa2012} and Rydberg atoms \cite{Saffman2010, Schausz2012} now offers the unique opportunity to explore the interplay between novel single-particle band structures and \textit{long range} interactions in systems that have no traditional solid state analogues. Here we study the simplest example of such a system, which can be readily realized in current experiments: a spinless dipolar Bose gas trapped on a two-leg ladder in a large magnetic field. 

 Motivated by ongoing experiments on polar molecules, Rydberg and magnetic atoms, the rich physics of dipolar gases in low dimensional systems has recently been explored by several authors: in arrays of one-dimensional ($1$D) optical lattices, at low densities, strong density-density interactions give rise to ordered crystalline phases at rational filling fractions forming a devil's staircase \cite{Dalmonte2010, Parish2012, Burnell2009}. By taking advantage of the low-lying rotational states of polar molecules, and the anisotropy of the dipolar interaction, spin Hamiltonians such as the  $\text{XXZ}$ spin chain with direction dependent couplings, or the bi-quadratic spin-$1$ Haldane Hamiltonian can be realized in deep lattices, allowing the study of symmetry protected topological phases on two leg ladders \cite{Manmana2013, Liu2012}. In engineered lattice potentials with non-trivial band topology, dipolar interactions naturally give rise to lattice analogues of the fractional quantum Hall effect, namely fractional Chern insulators \cite{Yao2013}. In the continuum, spin-orbit coupled bosons with long range interactions possess ground states with novel topological defects or quasi-crystalline order \cite{Sarang2013, Wilson2013}. Our study complements these earlier works by focussing on the weak coupling limit of large onsite occupation, where the interplay between kinetic frustration and long range interactions on a two leg ladder leads to superfluid phases with broken translational and reflection symmetries. 

 
 Experimentalists at Munich recently engineered a system of bosons on a two-leg ladder by using Raman lasers to create a uniform magnetic flux $\phi$ per plaquette \cite{Atala2014}.  They explored the non-interacting physics at a fixed flux $\phi = \pi/2$ and found two phases as a function of the rung-to-leg coupling strength: a saturated chiral current or Meissner phase at large coupling, where equal and opposite currents flow on each leg of the ladder, and a modulated density or vortex phase at small coupling, where the density is modulated along the legs of the ladder. These phases were first discussed theoretically by Orignac and Giamarchi \cite{Orignac2001} using bosonization, although their existence had already been predicted in the context of weakly coupled Josephson junction arrays \cite{Kardar1986}. The strongly interacting limit of this problem has now been thoroughly explored, where the interplay between single-particle degeneracies and interactions leads to interesting physics such as Mott phases with staggered loop currents \cite{Dhar2012, Dhar2013, Petrescu2013, Piraud2015, Greschner2015, Keles2015, Tokuno2014}. Following the experiment of Atala \textit{et al.} \cite{Atala2014}, Wei and Mueller \cite{Wei2014} used a variational approach to explore the effects of weak, short-range repulsive interactions, finding an additional phase, dubbed the biased ladder phase, where the density is uniform but different on each ladder, thereby breaking global $Z_{2}$ reflection symmetry. Here we generalize the theory and results of Ref.~\cite{Wei2014} to long range dipolar interactions, finding a rich phase diagram as a function of the dipolar interaction strength, the synthetic magnetic field, and the relative tilt angle between the external field polarizing the dipoles, and the plane of the ladder. 
 
Our main results are summarized below:
\begin{enumerate}

\item Generally long range dipolar interactions either destroy, or reduce the region of stability of the Meissner or saturated chiral current phase.

\item Repulsive dipolar interactions produce an interleg charge density wave (CDW) phase where the densities along the left and right legs of the ladder modulate out of phase with one another. 

\item Next nearest neighbor interactions support a fully modulated biased ladder phase, where all the particles are located on one leg of the ladder.

\item Arbitrarily weak attractive interactions along the rungs destroy the Meissner phase entirely, and lead to a cascade of first order transitions between distinct modulated density wave phases with different wave-vectors. 

\item Attractive nearest neighbor interactions along the ladder produce a regime of parameters where the biased ladder phase is the stable ground state at weak rung hopping. 

\end{enumerate}

\section{The Model}

\begin{figure}
\begin{picture}(100, 50)
\put(-88, -10){\includegraphics[scale=0.425]{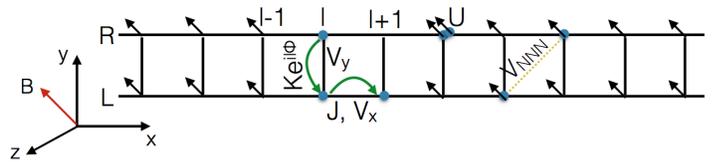}}
\end{picture}
\caption{\label{schematic} (Color Online) \textbf{Dipolar Bose gas on a two-leg ladder}. The arrows on the sites of the ladder correspond to dipoles, aligned in the direction of the external field indicated by $B$. The external field could be an electric field in the case of polar molecules or a magnetic field for magnetic atoms such as Dy and Er. The hopping along the $x$ direction is denoted by $J$. An artificial flux threads the system, which implies that hopping in the $y$ direction picks up a position dependent phase $e^{il\phi}$. The local on-site interaction is denoted by $U$ and is always assumed to be repulsive $U \geq 0$. Additionally, there is a nearest neighbor interaction along $x$ (ladder) and $y$ (rung) denoted $V_{x}$ and $V_{y}$ respectively, and a next nearest neighbor interaction $V_{\text{NNN}}$. The sign and magnitude of $V_{x}$, $V_{y}$ and $V_{\text{NNN}}$ can be tuned by tilting the external field relative to the plane of the ladder. Higher order contributions to the dipolar interaction are significantly smaller in magnitude, and do not have much effect on the overall phase diagrams we present.}
\end{figure}

Our Hamiltonian for a two-dimensional spinless Bose gas on a two-leg ladder with lattice spacing $a$, takes the form (see Fig.~\ref{schematic}):
\begin{equation}\label{ham0}
{\cal{H}} = {\cal{H}}_{0} + {\cal{H}}_{\text{int}}
\end{equation}

where ${\cal{H}}_{0}$ reads:
\begin{eqnarray}\label{spham}
{\cal{H}}_{0} = -J\sum_{l}(a^{\dagger}_{l+1,L}a_{l, L} + a^{\dagger}_{l+1, R}a_{l, R} + \text{h.c}) \\\nonumber -K\sum_{l}(a^{\dagger}_{l, L}a_{l, R}e^{-il\phi}+ \text{h.c})
\end{eqnarray}
where $a_{l,L}$ and $a_{l,R}$ are bosonic annihilation operators on the left (L) and right (R) legs of the ladder at position $l$. $J$ and $K$ denote the hopping matrix elements along the legs and rungs of the ladder respectively, and $\phi$ is the magnetic flux per plaquette \cite{Atala2014}. This single particle model was introduced by Atala \textit{et al.} \cite{Atala2014}. 
 
We further assume that there is an external (real) magnetic or electric field which polarizes the magnetic atoms/polar molecules, freezing out any internal degrees of freedom. As a result, the interaction Hamiltonian only includes density-density interactions and takes the form:
\begin{eqnarray}
{\cal{H}}_{\text{int}} = \frac{U}{2}\sum_{l, \mu}(a^{\dagger}_{l, \mu}a^{\dagger}_{l, \mu}a_{l, \mu}a_{l, \mu}) + \\\nonumber 
+ \frac{1}{2}\sum_{l, l^{'}, \mu, \mu^{'}}V_{l, l^{'}, \mu, \mu^{'}}a^{\dagger}_{l, \mu}a^{\dagger}_{l^{'}, \mu^{'}}a_{l^{'},\mu^{'}}a_{l, \mu}
\end{eqnarray}
where $\mu, \mu^{'} \in (L, R)$, and $U$ denotes the repulsive onsite interaction potential, which includes s-wave and dipolar contributions. The non-local part of the dipolar interaction given by $V_{l, l^{'}, \mu, \mu^{'}} = V(1 - 3 \cos^{2}\theta_{l, l^{'}, \mu, \mu^{'}})|\bm{r}|^{-3}$, where $V = D^{2}/a^{3}$ ($D$ is the dipole moment), couples both left and right legs of the ladder. Here we define $\theta_{l, l^{'}, \mu, \mu^{'}}$ to be the angle between the external polarizing field and the vector $\textbf{r}$ made by the lattice sites $\{l, \mu\}$ and $\{l^{'}, \mu^{'}\}$.


Although the non-local dipolar interaction couples sites arbitrarily far apart, we restrict our calculations to nearest neighbor (NN) and next-nearest neighbor (NNN) interactions as shown in Fig.~\ref{schematic}. Previous studies on dipolar bosons in optical lattices have shown that retaining higher order terms has little \textit{qualitative} effect on the overall phase diagram \cite{Danshita2009}, but modifies the precise location of the phase boundaries. 

The anisotropic nature of the dipolar interaction means that the interactions between sites on the same leg ($\sim  a^{\dagger}_{l, \mu}a^{\dagger}_{l+1, \mu}a_{l+1, \mu}a_{l, \mu}$, or the $x$-direction) and sites on opposite legs ($\sim a^{\dagger}_{l, L}a^{\dagger}_{l, R}a_{l, R}a_{l,L}$, or the $y$-direction), can in principle be different, and can be experimentally controlled by changing the alignment of the external field with respect to the plane of the ladder (see Fig.~\ref{schematic}). We therefore separately denote these interactions as $V_{x}$ and $V_{y}$, and denote the next-nearest neighbor interaction ($\sim a^{\dagger}_{l, L}a^{\dagger}_{l+1, R}a_{l+1, R}a_{l,L}$) as $V_{\text{NNN}}$. 

Generally, the on-site ($U$) and dipolar ($V$) interaction strengths are tunable by changing the transverse confinement \cite{Chinconfine}. The contact interaction is separately tunable using Feshbach resonances \cite{ChinRMP2010, Pfau2007}. We only consider repulsive on-site potentials in this work \cite{manpreet2014}. In experiments, the sign of the non-local part of the dipolar interaction $V$ is separately tunable from repulsive to partially attractive by tilting the external applied field at an arbitrary angle $(\theta, \chi)$ relative to the ladder plane. For a general polar angle $\theta$ and azimuthal angle $\chi$, the interaction potentials along the ladder and rung direction read: $V_{x} \propto V(1- 3\sin^{2}{\theta}\cos^{2}{\chi})$ and $V_{y} \propto V(1- 3\sin^{2}{\theta}\sin^{2}{\chi})$ \cite{Danshita2009}. A major advantage of this is that the physics of attractive ladders can be accessed, while ensuring overall mechanical stability from a repulsive local interaction \cite{manpreet2014}.

Completely attractive dipoles can also be obtained using a time-dependent external field \cite{Pfau2010}. In addition to the case where the external field is polarized perpendicular to the ladder, where all the interactions are repulsive: $V_{x} = V_{y} = V$, $V_{NNN} = V/2\sqrt{2}$, we separately present the general phase diagram for attractive rung ($V_{y} < 0$) and ladder interactions ($V_{x} < 0$) respectively. For convenience, we focus on two particular tilt configurations: (i) external field along the rung ($y$) direction. Here $V_{y} = -2V$, $V_{x} = V$, $V_{NNN} = -V/4\sqrt{2}$  and 
(ii) external field along the ladder direction. Here $V_{x} = -2V$, $V_{y} = V$ and $V_{NNN} = -V/4\sqrt{2}$. Our qualitative results however are general, and robust to tilting the external field away from these angles, as long as the sign of the interaction along the ladder or rung direction does not change. 

The single-particle Hamiltonian in momentum space reads ${\cal{H}}(k) = -2 J \cos k\cos \frac{\phi}{2} + 2 J \sin k\sin \frac{\phi}{2} \sigma_{z} - K \sigma_{x}$, and is readily diagonalized to yield two bands, whose energies are $E_{\pm} = -2 J \cos k\cos\phi/2 \pm \sqrt{4 J^2 \sin^{2} k \sin^{2} \phi/2 + K^{2}}$ \cite{Wei2014, Carr2006, Roux2007}. The lowest band has two minima at $k = \pm k_{0}$ for $K < 2 J \tan\phi/2\sin\phi/2$, and a single minimum at $k= 0$, for $K$ greater than this value. 

\subsection{Variational Approach I}

In this work, we consider two complimentary variational approaches, valid for weak interactions, to study the ground state phase diagram of Eq.~(\ref{ham0}). 
The first approach follows that of Wei and Mueller \cite{Wei2014}, who considered the variational ground state wave-function for $N$ particles, restricted to the lowest band:
\begin{equation}\label{gs}
|G_{k_{0}}\rangle = \frac{1}{\sqrt{N!}}(\cos \gamma \beta^{\dagger}_{k_{0}}+\sin \gamma \beta^{\dagger}_{-k_{0}})^{N}|0\rangle
\end{equation}
where $\beta_{\pm k_{0}}$ are the annihilation operators for bosons at $k = \pm k_{0}$. The original boson operators can be expressed in terms of $\beta_{k}$ as $a_{kL} = -\sin{\frac{\theta_{k}}{2}}\beta_{k}$ and $a_{kR} = \cos{\frac{\theta_{k}}{2}}\beta_{k}$, where the angle $\tan{\theta_{k}} = \frac{-K/J}{2\sin{k}\sin{\phi/2}}$. Here $|0\rangle$ denotes the vacuum state, and $0 < \gamma < \pi/2$ for $k_{0} >0$, and $\gamma = 0$ for $k_{0} = 0$. Absent interactions, this is the ground state for any $\gamma$, but arbitrarily weak interactions will break this infinite degeneracy. 

\subsection{Observables}

The local density on each leg is defined as $n_{l,\mu} = \langle G_{k_{0}}|a^{\dagger}_{l,\mu}a_{l,\mu}|G_{k_{0}}\rangle$. We define the average density as $n = N/\Omega$, where $\Omega$ is the volume of the system. The expressions for the average density on each leg and its modulations are derived in Ref.~\cite{Wei2014}, but reproduced here for completeness:
\begin{eqnarray}\label{densformulas}
n_{l,L} =  \cos^{2}{\gamma}\sin^{2}{\theta_{k_{0}}} + \sin^{2}{\gamma}\cos^{2}{\theta_{k_{0}}} + \\\nonumber \frac{1}{2}\sin{2\gamma}\sin{\theta_{k_{0}}}\cos(2k_{0}l) \\\nonumber
n_{l,R} =  \sin^{2}{\gamma}\sin^{2}{\theta_{k_{0}}} + \cos^{2}{\gamma}\cos^{2}{\theta_{k_{0}}} +\\\nonumber \frac{1}{2}\sin{2\gamma}\sin{\theta_{k_{0}}}\cos(2k_{0}l).
\end{eqnarray}

A key limitation of variational approach I is that the modulations on the $L$ and $R$ legs of the ladder are identical. While this is a valid approximation for short ranged interactions considered in Ref.~\cite{Wei2014}, where the interaction does not couple the two legs of the ladder, this is no longer the case for dipolar interactions. 


Working in momentum space, we define the current $j_{\mu} =  \langle G_{k_{0}}| \sum_{k} a^{\dagger}_{k, \mu}\frac{\partial {\cal{H}}}{\partial k} a_{k, \mu} |G_{k_{0}}\rangle$. The total current $j_{\text{net}} = \sum_{\mu}j_{\mu}$ is identically zero in equilibrium, but the chiral current $j_{c} = j_{L} - j_{R}$ can be finite. In order to visualize phases, it is also helpful to define the current in real space along the legs and rungs of the ladder as \cite{Piraud2015}:
\begin{eqnarray}\label{cureqs}
j^{\parallel}_{l, \mu} = i J(a^{\dagger}_{l+1, \mu}a_{l, \mu} -a^{\dagger}_{l, \mu}a_{l+1, \mu})\\\nonumber
j^{\perp}_{l} = i K(e^{-il\phi}a^{\dagger}_{l, L}a_{l, R} -e^{il\phi}a^{\dagger}_{l, R}a_{l, L})  
\end{eqnarray}
This is particularly useful for comparison with real space numerical methods such as DMRG. 
Explicit expressions for the rung and ladder currents in real space can be calculated using the variational wave-function Eq.~(\ref{gs}), and are provided in the Appendix. 
The net and chiral currents are defined as  $j_{\text{net}/c} = \sum_{l}(j^{\parallel}_{l, L} \pm j^{\parallel}_{i, R})$ respectively. 

Using variational approach I for short range interactions ($V = 0$), Wei and Mueller \cite{Wei2014}, found three phases: 
\begin{enumerate}
\item Meissner (or saturated chiral current) phase.  Here the rung current vanishes, and equal and opposite currents flow along the legs of the ladder. The total density is uniform. 
\item A biased ladder (BL) phase, where the density is uniform but different on the left and right legs, breaking global $Z_{2}$ reflection symmetry, in addition to the global $U(1)$ symmetry of the condensate. The rung current is zero. 
\item A modulated density (or vortex) wave phase, where the density is identical on the left and right legs but oscillates with a period incommensurate with the underlying lattice. This phase breaks additional $U(1)$ symmetry, associated with translations of the density wave. 
\end{enumerate} 

The presence of local currents on each plaquette in the modulated density phase are analogous to vortices in a type-II superconductor. The terminology of vortex and Meissner phases was first introduced by Orignac and Giamarchi \cite{Orignac2001}, and Atala \textit{et al.} \cite{Atala2014} interpreted their data using this language. Here we will interchangeably use the terminology saturated chiral current (or Meissner) and modulated density wave (or vortex) phase. We caution however that for strong interactions, vortex states may not always be accompanied by density modulations, indeed homogeneous insulating vortex phases have been found in DMRG studies \cite{Piraud2015}. As we explicitly derive in the Appendix, in all the three phases, the net ladder current $j_{\text{net}}$ remains zero on every plaquette. 

We note that in principle, one can construct general non-local density and bond operators such as $\langle a_{i\mu}a_{j\mu^{'}}\rangle$ for sites $i, j$ and $\mu, \mu^{'} \in \{L, R\}$ \cite{Roux2007, Carr2006}, which can take on nonzero values, especially for systems with long range interactions. In the weakly interacting limit however, whenever $\langle\beta_{k_{0}}\rangle$ becomes finite, these other order parameters automatically  take on non-zero expectation values. Therefore it suffices to only consider the density here. How these non-local order parameters vanish near the Mott transition however, is a separate question. 


The \textit{ansatz} of Eq.~(\ref{gs}) can be readily employed to calculate the total energy of the dipolar gas $E = E_{-}(k) + \langle G_{k_{0}}|{\cal{H}}_{\text{int}}|G_{k_{0}}\rangle$, where we treat $\gamma$ and $k_{0}$ as variational parameters. The saturated chiral current (biased ladder) phases correspond to $\gamma =0$ and $k =0 ~(\neq 0)$ respectively, whereas the vortex or modulated density phase corresponds to $\gamma \neq 0$, $k \neq 0$. A new feature of dipolar systems is that $\gamma$ can be pinned to a finite value, even as $k \rightarrow 0$, which implies a state with long wave-length density modulations, yet a fully saturated chiral current. This state breaks the same symmetries as the modulated density-wave phase. 

The variational form for the interaction energy is:
\begin{widetext}
\begin{eqnarray}\label{inte}
E(k_{0}, \gamma) = E_{-}(k_{0}) +\frac{(U+V_{x})n}{2}(1 - \frac{1}{2}\sin^{2}{\theta_{k_{0}}} -\frac{1}{2}\sin^{2}{2\gamma} + \frac{1}{2}\sin^{2}{2\gamma}\sin^{2}{\theta_{k_{0}}}) + \frac{(U + V_{x}\cos(2k_{0}))n}{8} \sin^{2}{2\gamma}\sin^{2}{\theta_{k_{0}}} \\\nonumber
+ \frac{(V_{y} + 2V_{\text{NNN}})n}{8}(\sin^{2}{\theta_{k_{0}}} + \sin^{2}{2\gamma} - \sin^{2}{\theta_{k_{0}}}\sin^{2}{2\gamma}) + \frac{(V_{y} + 2V_{\text{NNN}}\cos{(2k_{0})})n}{16}\sin^{2}{2\gamma}\sin^{2}{\theta_{k_{0}}}
\end{eqnarray}
\end{widetext}
One readily checks that in the absence of non-local dipolar interactions, $V_{x} = V_{y} = V_{\text{NNN}}= 0$, the interaction energy reduces to the form obtained in Ref.~\cite{Wei2014}. 

\subsection{Variational approach II}

It is useful to note that the three phases found in Ref.~\cite{Wei2014} are directly analogous to the corresponding phases of a weakly interacting spin-orbit coupled Bose gas in the continuum \cite{Paredes2014}, which has been extensively studied theoretically and experimentally  \cite{Li2012, Spielman2011, Ji2014, Zhang2011, Zhai2012, OurNature}. There, the one-dimensional spin-orbit coupling plays the role of the artificial magnetic field. The modulated density wave phase in the ladder problem corresponds to a stripe phase of the spin-orbit coupled Bose gas, where both single particle minima are occupied; the biased ladder corresponds to the situation where only one minimum is occupied, and the Meissner (saturated chiral current) phase corresponds to a non-magnetic phase, where $k_{0} = 0$. 

Motivated by this, we consider a second variational \textit{ansatz}, first introduced by Li \textit{et al.} \cite{Li2012} to study the spin-orbit coupled gas, but adapted to the present problem. This has the key advantage that it allows us to readily generalize to a situation where the modulations on the left and right legs of the ladder are unequal. We replace the boson operators on the left and right legs of the ladder by classical fields:
\begin{eqnarray}\label{ansatz}
a_{l, L} =e^{-\frac{i\phi l}{2}}\sqrt{\frac{N}{\Omega}}\Big(C_{1L}\sin{\frac{\theta_{k_{0}}}{2}}e^{ik_{0}l} + C_{2L}\cos{\frac{\theta_{k_{0}}}{2}}e^{-i k_{0}l}\Big)\hspace{3.5mm}\\\nonumber 
a_{l, R} = e^{\frac{i\phi l}{2}}\sqrt{\frac{N}{\Omega}}\Big(C_{1R}\cos{\frac{\theta_{k_{0}}}{2}}e^{ik_{0}l} + C_{2R}\sin{\frac{\theta_{k_{0}}}{2}}e^{-i k_{0}l}\Big) \hspace{5mm}
\end{eqnarray}
where $C_{1\mu},C_{2\mu}$ are complex numbers which satisfy the constraint: $\sin^{2}{\frac{\theta_{k_{0}}}{2}}(|C_{1L}|^{2} +|C_{2R}|^{2}) + \cos^{2}{\frac{\theta_{k_{0}}}{2}}(|C_{2L}|^{2} +|C_{1R}|^{2}) = 1$, from number conservation.  Note that the  variational approach of Refs. \cite{Wei2014, Li2012} corresponds to the limit where $C_{1L} = C_{1R}$ and $C_{2L} = C_{2R}$. This restricted \textit{ansatz} yields a density wave texture, but the spin density remains homogeneous. However, in the presence of inter-rung interactions, our more general \textit{ansatz} is naturally able to capture states with spin and density wave modulations, which are given by $n_{l} = n_{l, L} + n_{l, R}$ and $S_{l} = n_{l, R} - n_{l, L}$ where:

\begin{eqnarray}\label{densspinmod}
n_{l, L} = |C_{1L}|^{2}\sin^{2}{\frac{\theta_{k_{0}}}{2}} + |C_{2L}|^{2}\cos^{2}{\frac{\theta_{k_{0}}}{2}} + \\\nonumber \frac{1}{2}(C_{1L}C^{*}_{2L}e^{2ik_{0}l}\sin{\theta_{k_{0}}} + \text{c.c})\\\nonumber
n_{l, R} = |C_{2R}|^{2}\sin^{2}{\frac{\theta_{k_{0}}}{2}} + |C_{1R}|^{2}\cos^{2}{\frac{\theta_{k_{0}}}{2}} + \\\nonumber \frac{1}{2}(C_{1R}C^{*}_{2R}e^{2ik_{0}l}\sin{\theta_{k_{0}}} + \text{c.c})
\end{eqnarray}

Inserting Eq.~(\ref{ansatz}) into the Hamiltonian, the expression for the total energy reads:
\begin{widetext}\label{tote}
\begin{eqnarray}
E_{\text{MF}} = E_{-}(k_{0}) + \frac{(U+V_{x})n}{2}\Big((|C_{1L}|^{4} + |C_{2R}|^{4})\sin^{4}{\frac{\theta_{k_{0}}}{2}}+(|C_{1R}|^{4} + |C_{2L}|^{4})\cos^{4}{\frac{\theta_{k_{0}}}{2}}\Big)+ \\\nonumber \frac{(U+V_{x}\cos^{2}{k_{0}})n}{2}\sin^{2}{\theta_{k_{0}}}(|C_{1L}|^{2}|C_{2L}|^{2} +|C_{1R}|^{2}|C_{2R}|^{2}) + \frac{V_{y}n}{2}\Big(|C_{1L}|^{2}|C_{2R}|^{2}\sin^{4}{\frac{\theta_{k_{0}}}{2}} +|C_{2L}|^{2}|C_{1R}|^{2}\cos^{4}{\frac{\theta_{k_{0}}}{2}} + \\\nonumber \frac{1}{4}\sin^{2}{\theta_{k_{0}}}(|C_{1L}|^{2}|C_{1R}|^{2} + |C_{2L}|^{2}|C_{2R}|^{2} + C^{*}_{1L}C^{*}_{2R}C_{2L}C_{1R} + \text{c.c})\Big) + V_{\text{NNN}}n\Big(|C_{1R}|^{2}|C_{2L}|^{2}\cos^{4}{\frac{\theta_{k_{0}}}{2}} + \\\nonumber |C_{1L}|^{2}|C_{2R}|^{2}\sin^{4}{\frac{\theta_{k_{0}}}{2}} + \frac{1}{4}\sin^{2}{\theta_{k_{0}}}(|C_{1L}|^{2}|C_{1R}|^{2} +|C_{2L}|^{2}|C_{2R}|^{2} + C^{*}_{1L}C^{*}_{2R}C_{2L}C_{1R}\cos{2k_{0}} + \text{c.c})\Big)
\end{eqnarray}
\end{widetext}

Note that unlike the total energy of variational approach I, $E_{\text{MF}}$ is no longer symmetric under $k_{0} \rightarrow -k_{0}$. This is because under $k_{0} \rightarrow -k_{0}$, $\cos\frac{\theta_{-k_{0}}}{2} \rightarrow \sin\frac{\theta_{k_{0}}}{2}$. Therefore, for generic values of the co-efficients $C_{1\mu}$ and $C_{2\mu}$, $E_{\text{MF}}(k_{0}) \neq E_{\text{MF}}(-k_{0})$. One way to impose this symmetry is by setting $C_{1L} = C_{1R}$ and $C_{2L} = C_{2R}$ in the variational \textit{ansatz}. Here we look for more general solutions by numerically minimizing the variational energy $E_{\text{MF}}$ for arbitrary complex $C_{1\mu}$ and $C_{2\mu}$, with the constraints of particle number conservation and $E_{\text{MF}}(k_{0}) = E_{\text{MF}}(-k_{0})$. 


In what follows, we use both variational approaches to obtain the ground state energy. For repulsive interactions along the rungs, the interleg charge density wave solution found using variational approach II always has lower energy than the modulated density wave phase of variational approach I. All other phases we find are captured by both \textit{ansatzes}. Moreover, both approaches yield identical phase boundaries.   

Before turning to the results, we comment on the validity of variational, mean-field approaches in $1$D, where fluctuations are important and destroy long range order. The crucial parameter controlling the validity of our approximations is the ratio of the interaction to the kinetic energy $\zeta = E_{\text{int}}/E_{\text{kin}}$. In $1$D, the interaction energy scales as $E_{\text{int}} \sim Un$, while the kinetic energy scales as $E_{\text{kin}} \sim \hbar^{2}/2md^{2}$, where $d$ is the mean interparticle-spacing $d \propto 1/n$. The weakly interacting, mean-field regime occurs when the parameter $\zeta \propto U/n \ll 1$ \cite{Gangardt2003}. Thus our approximation works best at high densities, which corresponds to having a large number of bosons per site, which is precisely the regime we consider here.

\section{Purely Repulsive Dipoles}

\subsection{Qualitative Features}

We begin by examining the phase diagram of the two-leg ladder for purely repulsive dipoles. Before turning to the numerical results, we discuss the qualitative physics we expect from long range interactions. We first consider the role of the ladder and rung interactions separately, and then present the full phase diagram. 

\begin{figure*}
\begin{picture}(100, 150)
\put(120, 95){\includegraphics[scale=0.48]{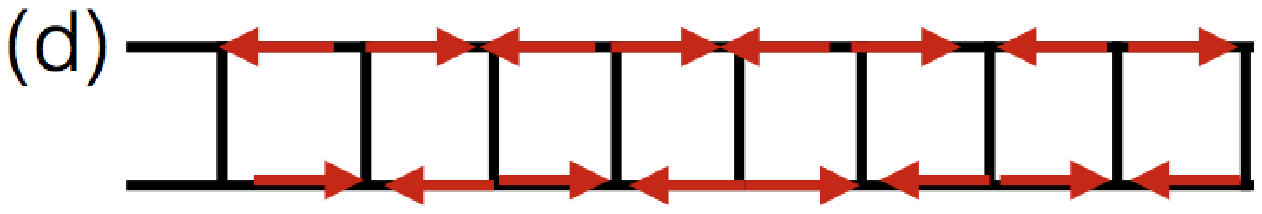}}
\put(135, -10){\includegraphics[scale=0.45]{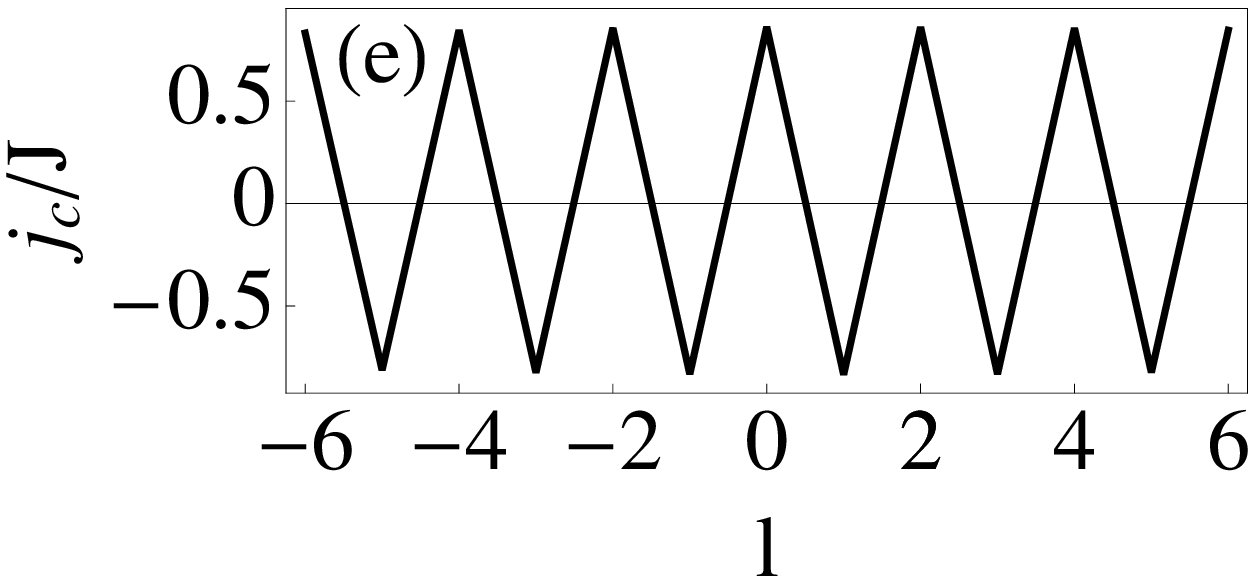}}
\put(-25, 70){\includegraphics[scale=0.4]{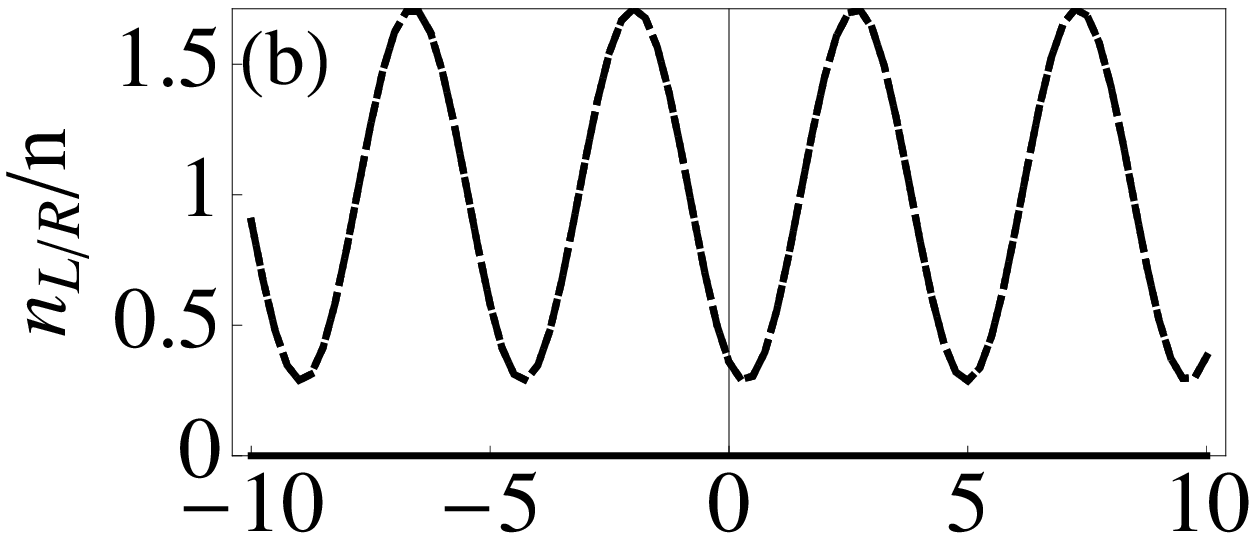}}
\put(-25, -10){\includegraphics[scale=0.4]{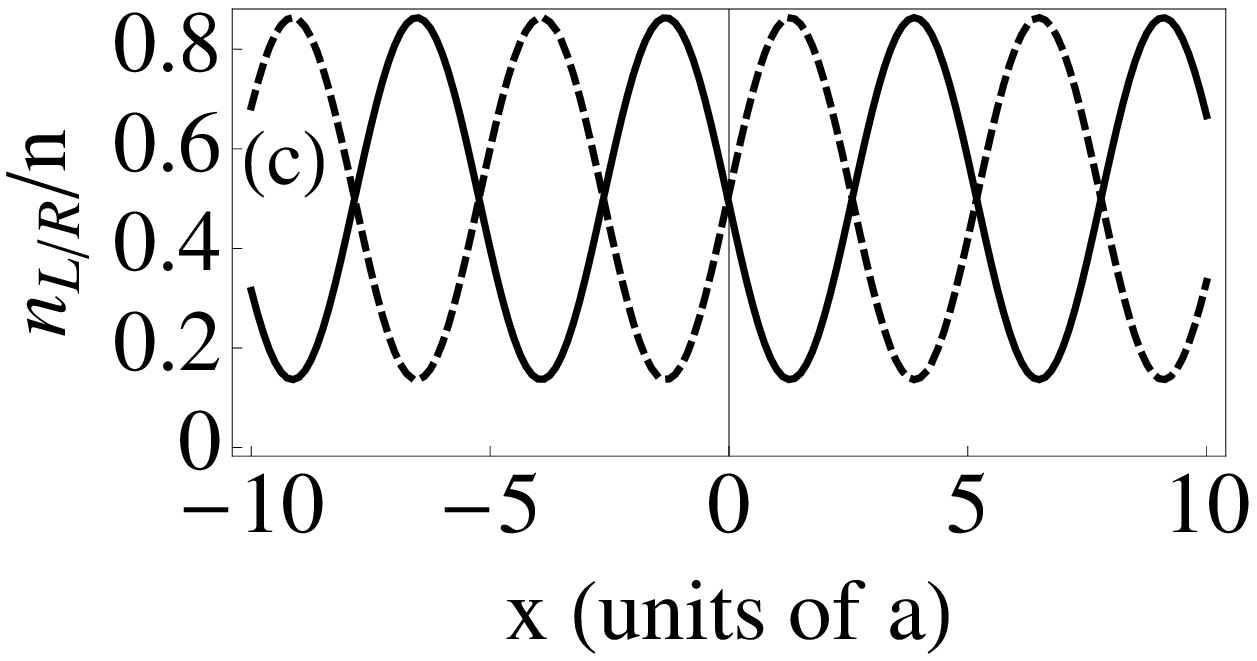}}
\put(-210, -10){\includegraphics[scale=0.42]{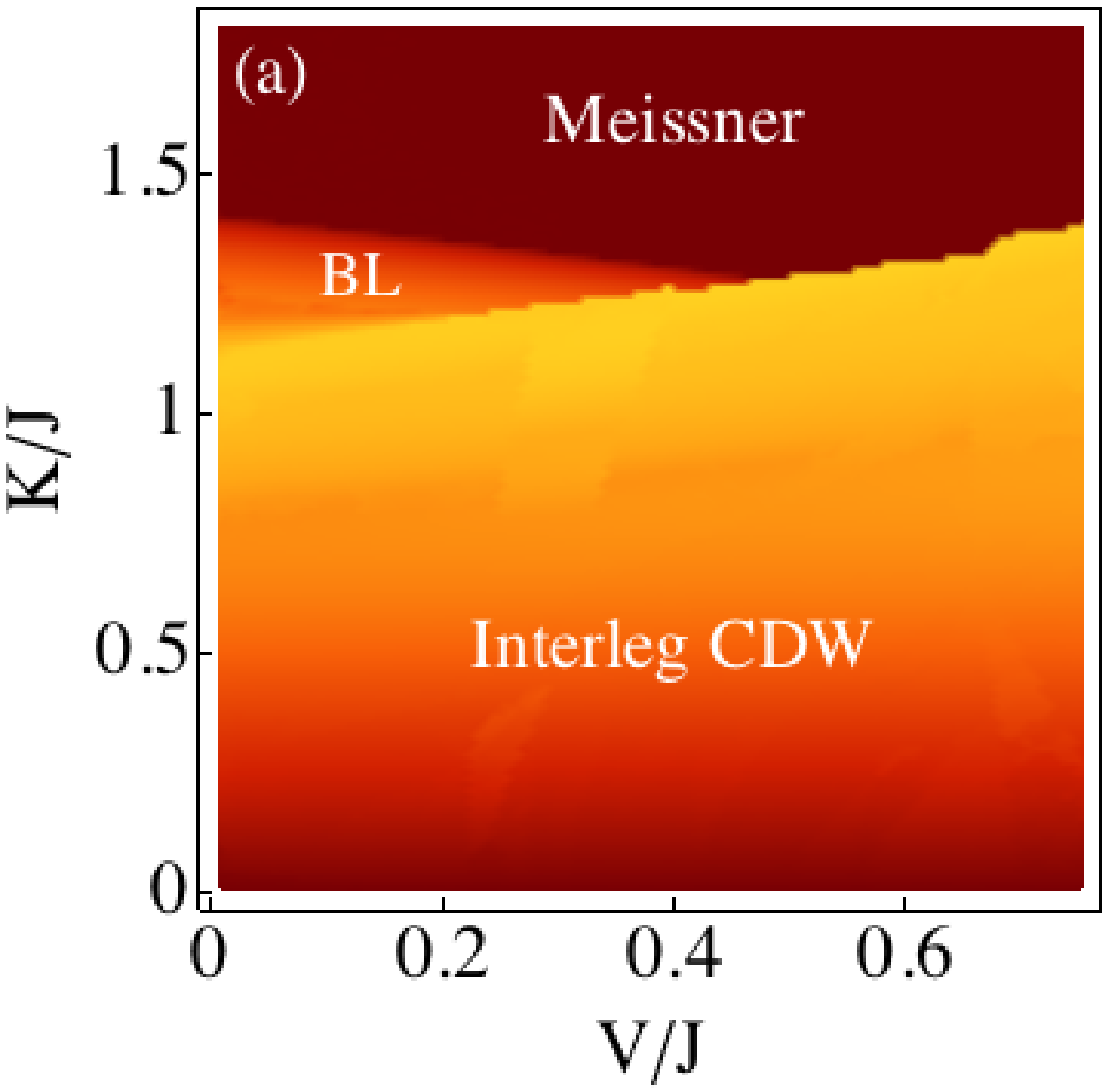}}
\put(-54, 15){\includegraphics[scale=0.45]{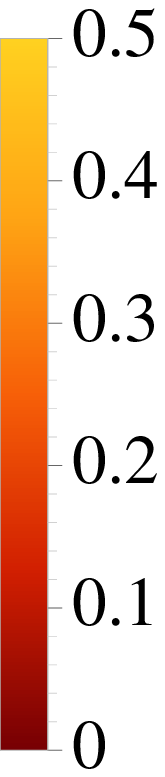}}

\end{picture}
\caption{\label{interlegcdw} (Color Online) \textbf{Repulsive Dipoles:} (a) Global phase diagram showing the density difference between left and right legs of the ladder as a function of $V/J$ and $K/J$ at $U = 0$ and $\phi = \pi/2$. Dipolar interactions push the Meissner phase to stronger rung-ladder coupling and give rise to an interleg CDW phase, where the relative densities on each leg of the ladder modulate out of phase with one another in real space as shown in (c). At intermediate rung hopping and weak dipolar interactions, a biased ladder (BL) phase is present. (b) Real space density profiles in the left (solid), right (dashed) and total density (dotted) for large next nearest neighbor interactions. All the particles reside on only one leg of the ladder, producing a fully modulated biased ladder phase. (d) Real space current profile at $\phi = 0.9\pi$, showing switching of the sign of the chiral current. Dhar \textit{et al.} \cite{Dhar2012} refer to this phase as a chiral superfluid. (e) Chiral current at $\phi = 0.9\pi$, $U=0$, $V/J = 1$ and $K/J = 1.5$.}
\end{figure*}

The nearest neighbor interaction ($V_{x}n_{l, \mu}n_{l+1\mu}$) along the ladder can be eliminated by placing particles on either the even or odd sites of the lattice, which corresponds to a modulated density wave phase. By contrast, it penalizes states with homogeneous density profiles on the legs of the ladder, which is the Meissner and biased ladder phases. We thus expect the dipolar ladder interaction to suppress these phases at large rung to ladder coupling strength ($K/J$). 

To see this, note that in Eq.~(\ref{inte}), this interaction contributes to the on-site interaction, and also yields a momentum dependent term term proportional to $\cos(2k_{0})$, which can compete with the local interaction $U$. In the following, we set $V_{\text{NNN}}$ and $V_{y}$ to zero in Eq.~(\ref{inte}) for simplicity. Although this choice is somewhat artificial, it is instructive in developing a systematic understanding about the effect of long range interactions. 

For purely contact interactions, when $K > K_{c}$, the single particle minimum occurs at $k=0$, and the energy of the Meissner state ($k=\gamma = 0$) reads: $E_{-}(0) + U/4 < E_{-}(k_{0}) + U/4 + U/8\sin^{2}{\theta_{k_{0}}}$, which is the energy of the modulated density state at $k = k_{0}$ and $\gamma = \pi/4$. Thus the Meissner phase wins at large $K$. By contrast, for nearest neighbor dipolar interactions along $x$, the energy of the modulated density state reads: $E_{-}(k_{0}) + V_{x}/4 + V_{x}\cos(2k_{0})\sin^{2}{\theta_{k_{0}}}/8$, which can be made lower than the Meissner phase energy ($E_{-}(0) + V_{x}/4$), by choosing $\pi/4 < k_{0} < \pi/2$, provided $V_{x}$ is sufficiently large. Therefore, for \textit{any} value of $K$, however large, there is always a transition out of the Meissner phase into a modulated density wave phase, provided $V_{x}$ is made sufficiently large. For typical parameters $K/J \sim 1.5$, the Meissner phase is fully destroyed in favor of a modulated density wave phase for $V_{x}/J \sim 7$. 


We now consider the effect of a repulsive inter-rung interaction on the phase diagram, which will naturally be present, whenever the external field is polarized perpendicular to the plane of the ladder. While this term also favors non-zero $k_{0}$ and $\gamma = \pi/4$, namely a modulated density phase, repulsive rung interactions penalize the left and right rungs from having identical density modulations. Indeed, we would expect that strong inter-rung repulsion would lead to ``phase separation", where the density modulations on the left and right legs are \textit{out-of-phase} with one another. We term such a state an \textit{interleg charge density wave} (CDW). 
This state therefore breaks local $Z_{2}$ reflection symmetry, even though globally, $Z_{2}$ reflection symmetry is unbroken, as the densities oscillate about the same average value. In the absence of gauge fields, such a state would be the ladder analog of a checkerboard solid phase of dipolar bosons in optical lattices \cite{Danshita2009, Goral2002}. However, our state additionally has non-zero rung and ladder currents arising from the synthetic flux threading the ladder. We remark that the interleg CDW phase we find here is the weak coupling analog of the ``SDW" phase found by Petrescu and Le Hur \cite{Petrescu2013} in the strong coupling limit of our model (absent next nearest neighbor interactions) at unit filling fraction. Furthermore, unlike the checkerboard phase of dipolar lattice bosons or the SDW phase, the interleg CDW phase we find is generally incommensurate with the underlying lattice, as the wave-vector of the density wave oscillations is set by the minima in the single particle dispersion.



\subsection{Interleg CDW}

The global phase diagram found by numerically minimizing the total energy $E_{\text{MF}}$ (Eq.~10) for $V_{x} = V_{y} = V$ and $U = V_{\text{NNN}} = 0$, and $\phi = \pi/2$, is shown in Fig.~\ref{interlegcdw}(a). Indeed for repulsive dipolar interactions, we find that a large portion of the phase diagram is occupied by the interleg charge density wave phase. From Eqs.~(\ref{densspinmod}), we plot the densities on the left (solid) and right (dashed) legs of the ladder, which are oscillating out of phase with one another. 

As expected from our discussion above, the Meissner phase indeed gets pushed to larger $K/J$ with increasing dipolar interactions. We have checked that variational approach I yields the same qualitative phase diagram and identical phase boundaries, but that theory cannot distinguish between a modulated density wave phase and an interleg CDW phase. The latter always has lower energy for repulsive rung interactions. For even larger $V$ (not shown), the Meissner phase disappears completely, as we argued from our qualitative discussion above. The transition from the interleg CDW to the Meissner phase is always first order for repulsive dipoles. For intermediate $K$, and weak dipolar interactions there is a biased ladder phase, which spontaneously breaks global $Z_{2}$ reflection symmetry, as the average densities on the left and right legs are no longer equal. For very weak dipolar interactions, our numerics find a \textit{modulated} biased ladder phase, where the densities on each leg modulate in real space but about different average values. However, if such a phase exists, we believe that it should only be present in a small window of the phase diagram. 

Inclusion of the next nearest neighbor term $V_{\text{NNN}}$, has no qualitative effect on the phase diagram presented above, as the next nearest neighbor interactions are significantly weaker than the ladder and rung interactions. Quantitatively however, next nearest neighbor interactions increase the window of stability of the biased ladder phase. Interestingly, if the next nearest neighbor interaction is the largest interaction in the problem, then the ground state is a \textit{fully} modulated biased ladder, where all the bosons reside on only one leg of the ladder (see Fig.~\ref{interlegcdw}(b)). This state not only breaks global $Z_{2}$ reflection symmetry, but also has an additional Goldstone mode associated with breaking translational symmetry. This state is the ladder analog of the stripe phase of two-dimensional lattice dipolar bosons. Although such a phase is theoretically interesting, this parameter regime cannot be attained with dipolar interactions alone. 

Tilting the dipoles away from the plane perpendicular to the ladder does not affect any of the qualitative conclusions of Fig.~\ref{interlegcdw}, provided the nearest and next nearest neighbor interactions remain repulsive. The precise locations of the phase boundaries however will shift. 
 
Dipolar systems have long been sought after as ideal candidate systems for exploring ``supersolidity": a phase of matter which exhibits dissipationless flow analogous to a superfluid while simultaneously possessing crystalline order \cite{Lahaye2009}. However a major experimental challenge in observing supersolidity in dipolar systems is that the dipolar interaction strengths in most magnetic atoms and polar molecules that have been cooled to date are far too weak, and are easily overwhelmed by the contact interaction, which produces a homogeneous superfluid \cite{Danshita2009}. However, in the present system, translational symmetry breaking is not an interaction effect, rather a single particle one, owing to the multiple minima in the single particle dispersion. Having broken translational symmetry to produce a modulated density wave phase, there is no additional on-site energy cost to displacing the density modulations on the left and right legs of the ladder. But this lowers the overall dipolar energy by avoiding the rung interaction. Therefore a checkerboard state or an interleg CDW phase can occur for very weak dipolar strengths in this system, and survives even in the presence of a non-zero $U$.


\subsection{Chiral current switching}

We briefly discuss what happens at other values of the flux. Although the experiments do not tune the flux directly, DMRG studies have extensively explored the phase diagram for different values of the magnetic field \cite{Greschner2015, Piraud2015}. Remarkably, we find that near (but not equal to) $\phi = \pi$, the sign of the chiral current develops a switching pattern from plaquette to plaquette as shown in Fig.~\ref{interlegcdw}((d) and (e)). In Fig.~\ref{interlegcdw}(e), we plot the chiral current at $\phi = 0.9\pi$, which shows this sign reversal. A similar pattern is predicted to occur in the fully frustrated Bose Hubbard model studied by Dhar \textit{et al.} \cite{Dhar2012, Dhar2013}, who refer to this state as a chiral superfluid. We remark that this is different from the findings of Greschner \textit{et al.} \cite{Greschner2015}, where the global chiral current reverses sign compared to the chiral current at small fluxes. Here we find that the sign of the global chiral current is still the same as that for small fluxes. We find that this switching pattern occurs even for weak interactions, at large enough values of $\phi$, when the Meissner phase is destroyed in favor of a vortex phase. Increasing $K$ increases the magnitude of the chiral current at fixed flux, and should make this effect experimentally observable. 

\section{Partially Attractive Dipoles}

\begin{figure*}
\begin{picture}(100, 150)
\put(135, 70){\includegraphics[scale=0.42]{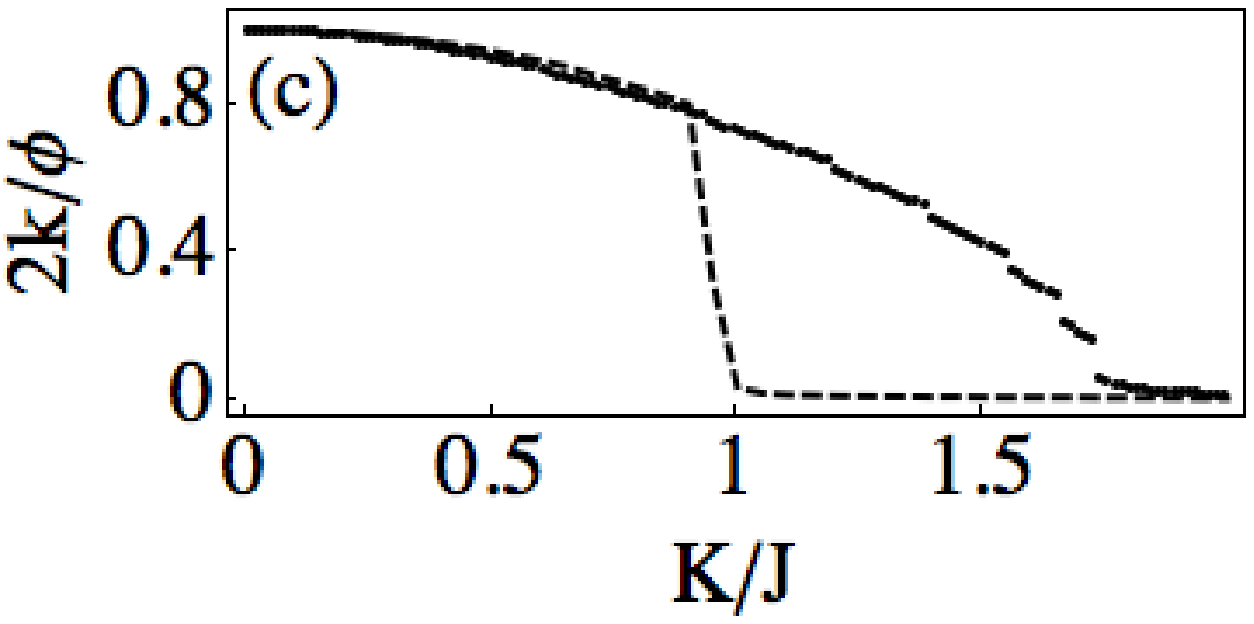}}
\put(-200, -5){\includegraphics[scale=0.42]{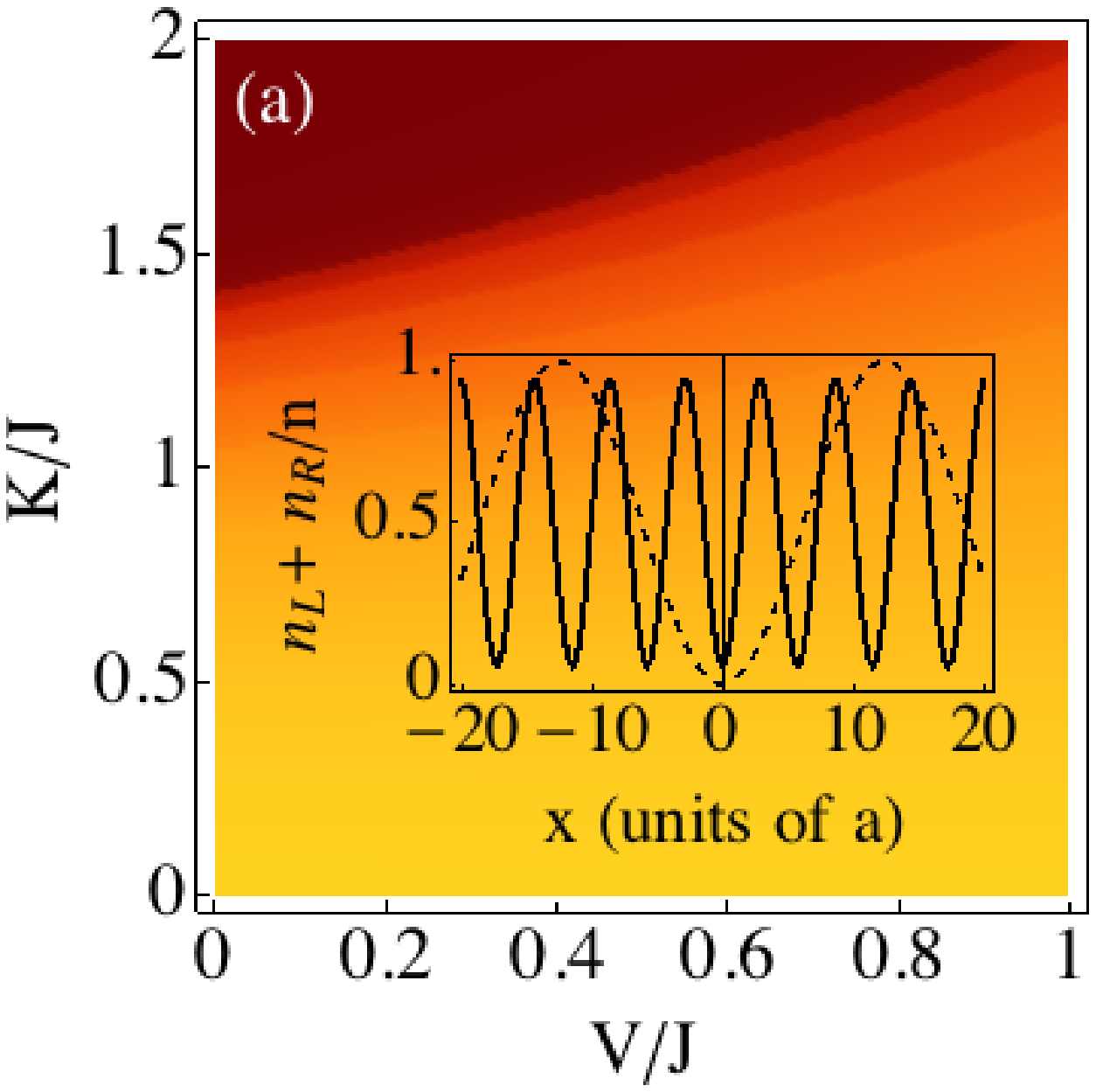}}
\put(-20, 70){\includegraphics[scale=0.42]{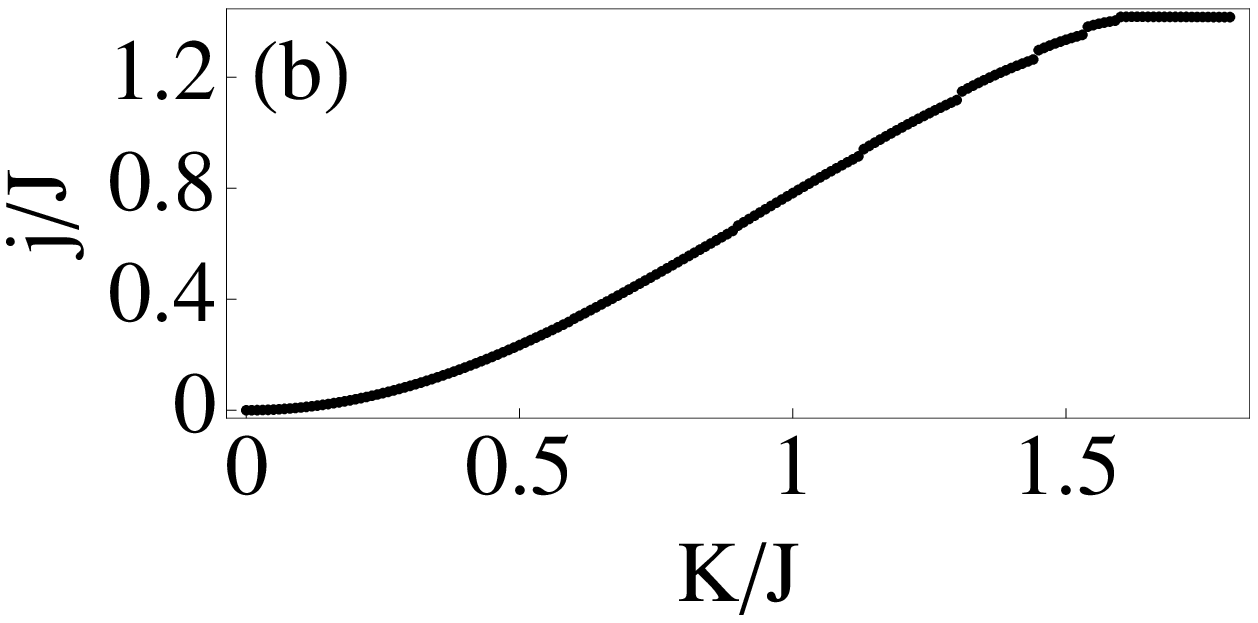}}
\put(40, 35){\includegraphics[scale=0.4]{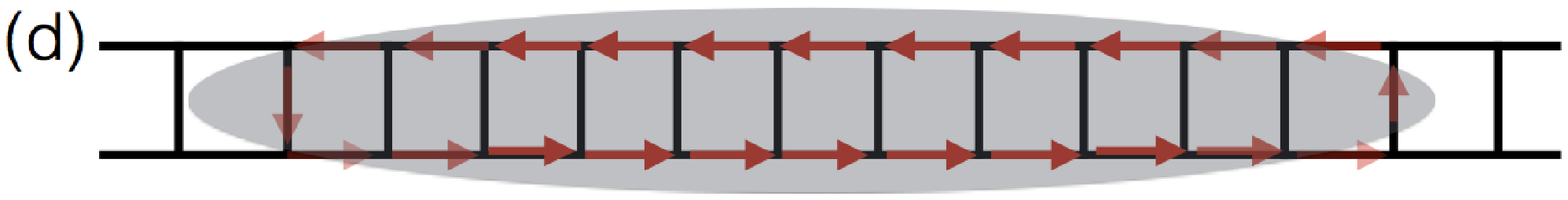}}
\put(-40, 20){\includegraphics[scale=0.42]{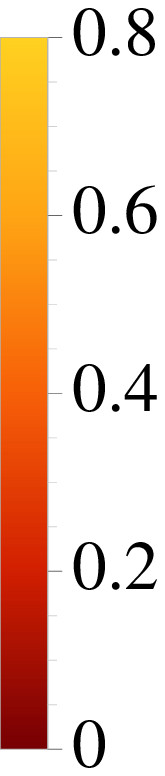}}
\put(30, 0){\includegraphics[scale=0.4]{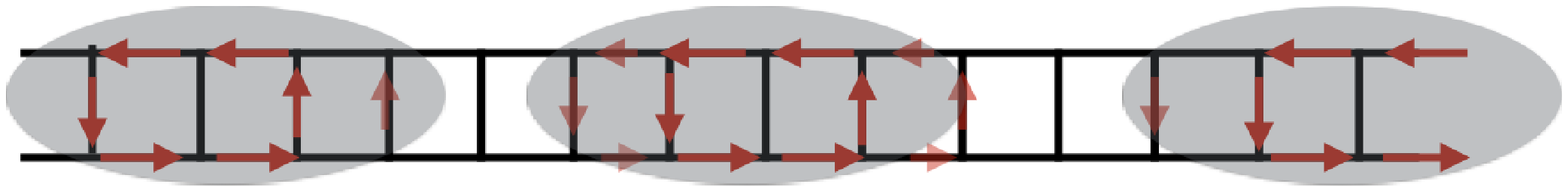}}
\end{picture}
\caption{\label{pdtiltrung} (Color Online) \textbf{Dipoles pointing along $y$ (rung) direction:} (a):  Phase diagram of the attractive rung, dipolar bosonic ladder for $U=0$, and $\phi = \pi/2$ as a function of $V$ and $K/J$. A modulated density wave phase (with $\gamma = \pi/4$) is found throughout the entire phase diagram. Density plot shows evolution of $k_{0}$ with $V/J$ and $K/J$, which approaches zero as $K$ becomes large, but $\gamma$ remains finite. Inset shows real space total density at two different values of $K$, at fixed $V$. ((b) and (c): Evolution of the chiral current and $k_{0}$ respectively for fixed $V$, as a function of $K/J$, showing jumps. The discrete jumps indicate first order transitions between CDW phases with different incommensurate wavelengths and is reminiscent of the devil's staircase pattern found in dipolar Mott insulators  and superfluids in $1$D \cite{Burnell2009, Dalmonte2010}. For comparison the evolution of $k$ for purely short range interactions is also shown (dashed), revealing a single first order transition from the modulated density phase to the Meissner phase. (d) Schematic showing the real space density and current profile in the two-leg ladder. Increasing $K/J$ increases the wavelength of the modulations (shaded region). Arrows indicate the direction of the chiral current In this picture, ``vortices" sit in the regions between the shaded areas, where the chiral current reverses direction. Therefore, the vortex density decreases as the $K/J$ increases (c.f. Ref.~\cite{Piraud2015} Fig. $5$). Both variational approaches never find a transition to a homogeneous density (Meissner) phase at any $K$.} 
\end{figure*}

We now turn to the study of partially attractive dipoles, which can be realized by tilting the external field relative to the plane of the ladder. For concreteness,  we focus on two particular cases: (i) external field aligned parallel to the rungs (attractive rung and next nearest neighbor interaction) and (ii) external field aligned parallel to the legs of the ladder (attractive ladder and next nearest neighbor interactions). We emphasize that even though we focus on specific tilt angles below, our results in Sec IV A and B are valid over a wide range of tilt angles as long as $V_{y} <0, V_{x} > 0$ and $V_{x} <0, V_{y} > 0$ respectively. As before we calculate the phase diagram using both variational approaches. Where comparison is possible, they yield identical results. 

\subsection{Attractive Rung interaction}

We first consider the case where the rung interaction $V_{y}$ is attractive and equals $V_{y} = -2V$, while the interaction along the ladder remains repulsive $V_{x} = V$. We also turn on a finite, attractive NNN interaction. As in the repulsive case, we consider a purely dipolar interaction $U=0$. Owing to the attractive rung interactions, the density modulations on both legs are in phase with one another, and both variational approaches yield a modulated density wave phase as the ground state. 

In Fig.~\ref{pdtiltrung}(a), we plot the phase diagram of the two leg ladder, as a function of $K/J$ and $V/J$ obtained from Eq.~(\ref{inte}). The density plot reveals that a vortex or modulated density phase ($\gamma = \pi/4$) occurs throughout the entire phase diagram, but that the wave-length of the modulations (or the vortex core radius) diverges as $K/J$ is increased. At large $K/J$, the chiral current saturates to its maximal value ($k_{0} \rightarrow 0$), but $\gamma$ remains pinned to $\pi/4$, so the density modulates in space. Thus we conclude that arbitrarily weak attractive rung interactions destroy the Meissner phase.

To understand the destruction of the Meissner phase, we once again consider the energy difference between the Meissner and a modulated density wave phase. Setting $V_{\text{NNN}} = 0$, which has no qualitative impact on the physics, the energy of the Meissner phase from Eq.~(\ref{inte}) reads: $E = E_{-}(0) -V/8$ while that of the modulated density wave phase is $E_{-}(k_{0}) - V/8 + V\sin^{2}({\theta_{k_{0}}})/8(1/2 \cos(2k_{0}) - 1)$. At large $K$, where the Meissner phase usually occurs, the energy difference between $E_{-}(0) - E_{-}(k_{0})$ approaches a constant independent of $K$, $E_{-}(0) - E_{-}(k_{0})\rightarrow -\sqrt{2}(1- \cos(k_{0}))$ . Therefore, for a modulated density wave phase to win, we require: $V\sin^{2}({\theta_{k_{0}}})/8(1/2~\cos(2k_{0}) - 1) <  -\sqrt{2}(1- \cos(k_{0}))$. It is easy to see that this condition can be satisfied even for arbitrarily weak $V$, as long as $k_{0} \rightarrow 0$. Therefore, for any value of $V$, there is a critical $k_{\text{crit}}$, such that for $k_{0}< k_{\text{crit}}$, the modulated density wave phase with wave-vector $k_{0}$ and $\gamma = \pi/4$ has lower energy than the Meissner phase. Hence the ground state is always a modulated density wave or vortex phase.

Turning on a finite $U$ however, restores the Meissner phase. Indeed, it is easy to show from Eq.~(\ref{inte}), that for $k_{0} \rightarrow 0$, a finite $\gamma$ occurs only when $U < U_{c} = V/4\sqrt{2}$. This condition can achieved experimentally by using Feshbach resonances to tune $U$ near a zero crossing. We expect that including higher order terms will increase $U_{c}$, as longer ranged interactions will be more attractive. However, we expect the quantitative corrections to be small, as longer range interactions decrease in magnitude as $1/\textbf{r}^{3}$. 

Physically the destruction of the Meissner phase for attractive rung interactions can be simply understood: Absent short-range repulsive forces ($U=0$), the dipolar interaction favors the formation of a CDW, and the attractive rung interaction implies that the CDWs on both legs will be in-phase with one another, \textit{i.e.}, a modulated density wave phase. Upon increasing $K$, the two single particle minima at $\pm k_{0}$ move closer to one another, but the total dipolar interaction energy can be lowered by condensing at \textit{both} minima, rather than only in a single minimum, which is the case for repulsive short-range interactions, which favors a homogeneous density. When $U$ becomes larger than a critical strength, the homogeneous density Meissner phase is restored.

Strikingly, the density plot reveals that $k_{0}$ shows a sequence of jumps as it varies from $k_{0} = \pi/4$ to $0$ with increasing $K$ (Fig.~\ref{pdtiltrung}(c)).  These jumps are also manifested in the chiral current (Fig.~\ref{pdtiltrung}(b)), and are indicative of a cascade of first order transitions between modulated density wave phases with different wave-vectors. The series of first order transitions between different modulated density wave phases are reminiscent of the Luttinger staircase feature observed in quasi-$1$D dipolar superfluids \cite{Dalmonte2010}. There, the dipolar Luttinger liquid becomes unstable towards the formation of a cascade of solids with fractional filling, in the presence of an arbitrarily weak lattice potential. Although we work in the opposite limit (namely of large incommensurate occupation), our theory nonetheless yields a similar cascade of CDW phases, where the periodicity of the density wave is tuned not by filling fraction, but rather by the hopping parameter. For comparison, in Fig.~\ref{pdtiltrung}(c) we also show the evolution of $k_{0}$ for a gas with contact interactions $U$ of the same magnitude, which shows a single first order transition from a modulated density wave to a Meissner phase. 



\subsection{Attractive Ladder interaction}

\begin{figure}
\begin{picture}(100, 160)
\put(-40, -10){\includegraphics[scale=0.49]{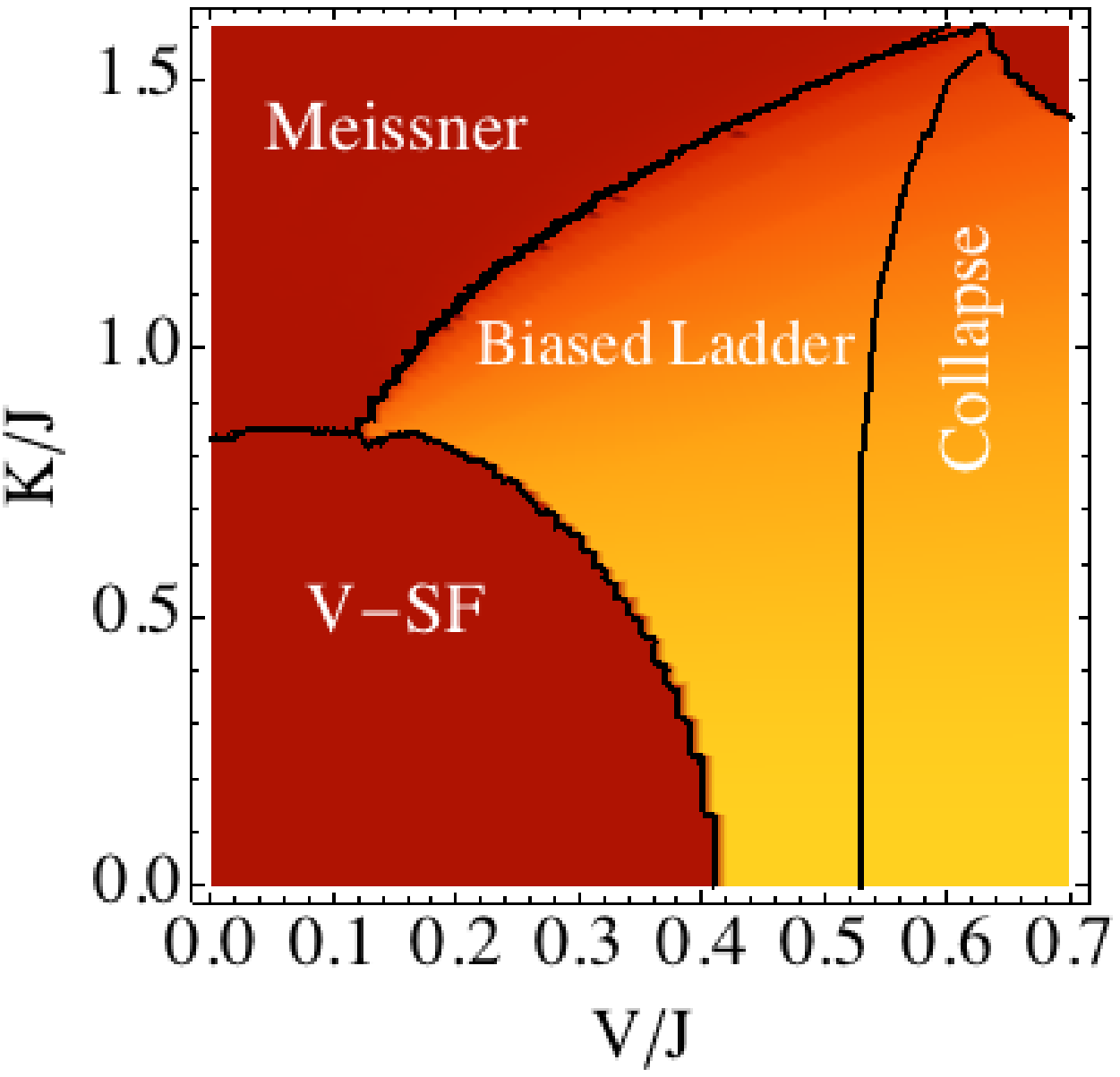}}
\put(140, 20){\includegraphics[scale=0.49]{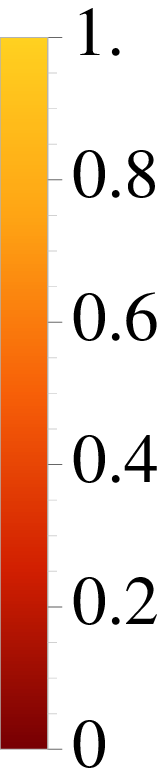}}
\end{picture}
\caption{\label{pdtiltladder} (Color Online) \textbf{Dipoles pointing along x (ladder) direction:} Density plot showing the density difference between the two legs of the ladder normalized to the total density for attractive ladder interactions: $V_{x} = -2V$, $V_{y} = V$, and $U/J = 1$. For finite $V/J$, there is a transition from an interleg CDW to a biased ladder phase at arbitrarily small $K/J$, which occupies a large portion of the phase diagram. Increasing $K/J$ leads to a direct transition from the biased ladder to the Meissner phase. Increasing the dipolar interactions further leads to a collapse of the gas, signaled by the appearance of imaginary frequencies at long wave-lengths.} 
\end{figure}

We now turn to the case where the magnetic field is tilted along the ladder ($x$) direction. We therefore have $V_{x} = -2V$, $V_{y} = V$. Numerically minimizing the variational energy Eq.~(10), the ground phase diagram for this case is shown in Fig.~\ref{pdtiltladder}. We have set $U/J = 1$. 

For weak ladder interactions $V << U$, the phase diagram resembles that of a gas with local interactions: a direct first order transition from an interleg CDW phase to a Meissner phase is found. (Note that for the choice of $U$, there is no intervening biased ladder phase (Ref.~\cite{Wei2014}).) However, increasing attractive dipolar interactions destroys the interleg CDW phase in favor of a biased ladder phase for arbitrarily weak rung to ladder couplings. Increasing $K/J$ leads to a second order transition from the biased ladder phase to the Meissner phase. We remark that the biased ladder phase has not yet been observed experimentally, despite being predicted theoretically \cite{Wei2014, Uchino2015, Piraud2015}. The tilted dipolar system displays a wide biased ladder regime, and may be ideally suited for observing this phase experimentally. 

A similar phase diagram, where the interleg CDW phase is replaced by a modulated density wave phase is expected for purely attractive short range interactions $U<0$. However, such a gas is not mechanically stable at zero temperature, and collapses at sufficiently high densities \cite{Mueller2000, Hulet1998}. In our case, the appearance of a biased ladder phase at small values of the rung hopping is driven entirely by attractive long range forces, and the gas can still be stabilized by maintaining a repulsive on-site potential. 

To compute the stability of the biased ladder phase, we calculate the excitation spectrum about the biased ladder ground state $|G_{k_{0}}\rangle =  1/\sqrt{N!}(\beta^{\dagger}_{k_{0}})^{N}|0\rangle$. Writing fluctuations above the ground state as $\beta_{k} = \sqrt{N}\delta_{k, k_{0}} + (1-\delta_{k, k_{0}})\gamma_{k-k_{0}}$ \cite{Wei2014}, we expand the Hamiltonian Eq.~(\ref{ham0}) to quadratic order in the $\gamma$ operators. Diagonalizing the effective Bogoliubov Hamiltonian $H - \mu N$, where $\mu = E_{-}(k_{0}) + (U+V_{x})n(\sin^{4}{(\theta_{k_{0}}/2)}+\cos^{4}{(\theta_{k_{0}}/2})) + (V_{y} + V_{\text{NNN}}/\sqrt{2})n\sin^{2}{(\theta_{k_{0}}/2)}\cos^{2}{(\theta_{k_{0}}/2)}$ is the chemical potential, yields the low energy excitation spectrum. As in the case of short-range interactions, the spectrum has a characteristic roton-maxon feature \cite{Wei2014}, which has also been recently found in the shaken lattice experiment of Parker \textit{et al.} \cite{Chin2013}. The dipolar interaction pushes the roton minimum to larger $k$. Near the interleg CDW to biased ladder phase transition, the roton minimum is present for arbitrarily small $K$, and disappears as $K$ is increased. 

The instability towards collapse is signaled by the appearance of imaginary frequencies at long wave-lengths. For $U/J = 1$, the gas becomes dynamically unstable towards collapse at $V/J \sim 0.5$, or $V_{x} \sim -U$. The critical interaction strength at which collapse occurs is weakly dependent on $K$. Increasing $K/J$ pushes the collapse to somewhat stronger interactions.


\section{Summary and Discussion}

Experimentally, our system can be realized using highly magnetic dipolar atoms, Rydberg atoms or polar molecules \cite{Ni2008, Aikawa2010, Deiglmayr2008,Lu2012, Pfau2007, Aikawa2012,Saffman2010, Schausz2012}. Magnetic atoms are ideally suited for this study, as we do not incorporate spin degrees of freedom, and therefore do not have any dipolar loss. Moreover, coupling magnetic atoms to Raman lasers is believed to have much lower heating losses \cite{Cui2013}, which is a major advantage over conventional alkalis. With polar molecules, a key experimental challenge that needs to be overcome is the low densities needed to prevent chemical reactions. Non-reactive molecules or quantum-Zeno methods of loss suppression \cite{Zhu2014}, may be required in order to boost the densities to values suitable for studying many-body physics.  

All the phases discussed here can be probed experimentally by measuring the \textit{in situ} density and the local currents \cite{Atala2014}. The interleg charge density wave phase breaks local $Z_{2}$ symmetry spontaneously. However, as the dipolar interaction is long ranged, even if there is no hopping between neighboring ladders, long range interactions will couple neighboring ladders, giving rise to an overall checkerboard pattern, which can be detected using Bragg spectroscopy \cite{Stenger99}. The modulated biased ladder phase is more challenging to detect, as the density asymmetry between the left and right legs of the ladder will randomly change from ladder to ladder, averaging to zero. However, by detuning the Raman lasers from resonance, an external bias can be applied \cite{Wei2014}, which then polarizes all the ladders in the same way. 

We comment briefly on the relationship between our work and recent experiments on bosons and fermions in synthetic dimensions \cite{Fallani15, Stuhl15}. The physical rung in the ladder we consider, corresponds to the synthetic dimension, implied by the internal spin states of the atoms. Therefore the onsite density-density and spin-spin interactions can be interpreted as effective ``long range" interactions in the synthetic (rung) dimension. A crucial difference however is that in these experiments, long range interactions only occur in the synthetic direction; in the physical direction, there is a one-dimensional optical lattice and the interactions are still short-ranged. If the analog of an interleg CDW occurs in these experiments, it will be manifest as a spin-density wave, where the local spin orientation changes from site to site, such that on a given rung of the ladder, only one spin component is present. 

To conclude, we have used two complementary mean-field approaches to explore the interplay between large magnetic fields and long range density-density interactions on a two-leg bosonic ladder, finding a rich phase diagram. Our approximations are justified in the large density limit of large numbers of bosons per site, which for $1$D systems corresponds to the mean-field regime. As in the experiment of Atala \textit{et al.} \cite{Atala2014}, we largely fix the magnetic flux, but modulate the rung-to-ladder hopping and the interactions. Generally we have shown that dipolar interactions destroy the Meissner phase completely or reduce its regime of stability. We expect that more sophisticated bosonization treatments should yield similar qualitative answers, as it is already well known that dipolar Luttinger liquids are unstable towards CDW formation \cite{Dalmonte2010}. A full bosonization treatment of this problem remains an exciting topic for further study. Repulsive density-density dipolar interactions lead to an interleg CDW phase, where the total density is uniform, but the densities along the left and right legs of the ladder modulate in space, out of phase with one another. This phase is stable for weak next nearest neighbor interactions, but for strong next nearest neighbor repulsion, a fully modulated biased ladder phase is found, where all the atoms are either on the left or the right leg of the ladder. For values of the flux near $\phi \rightarrow \pi$, we obtain a modulated density wave phase where the chiral current switches sign from plaquette to plaquette.

For purely attractive rung interactions, we find that the Meissner phase is completely destroyed. Instead, we find a cascade of first order transitions between CDW phases with different wave-vectors, reminiscent of the Luttinger staircase \cite{Dalmonte2010}. Whether this pattern develops into a full Luttinger staircase at low filling fractions will be studied in future work. Importantly, as these jumps can be tuned by changing the Raman coupling, this may open the possibility of observing this staircase pattern in experiments.  

Finally, we emphasize that although we have only considered specific tilt angles in this work for simplicity, our results are more general in that the phase diagrams obtained here will survive small deviations away from the tilt angles we consider. For example, the biased ladder phase is the ground state as long as $V_{x} <0$ and $|V_{x}| <U$ and $V_{y} >0$. This condition is met for all angles $\sin^{-1}{1/\sqrt{3}} < \theta < \pi/2$. Similarly, we expect an interleg CDW ground state for small $K/J$ whenever $V_{y} > 0$.

\section{Acknowledgements}
We are indebted to Ana Maria Rey for motivating us to think about this problem, and Wilbur Shirley and Marie Piraud for numerous discussions during the preparation of this manuscript. We are grateful to Xiaopeng Li, Erich Mueller, Ran Wei and Ryan Wilson for their careful reading of this manuscript and for suggesting numerous improvements. We would like to thank the LPS-CMTC, LPS-MPO-CMTC, NSF-JQI-PFC, and ARO-MURI and the NSF-PFC seed grant ``Emergent phenomena in interacting
spin-orbit coupled gases" for support. We are grateful to the Department of Energy's Institute for Nuclear Theory at the University of Washington for its hospitality, during the completion of this work. 

\begin{appendix}

\section{Expressions for local ladder and rung currents}

To calculate the currents in Eq.~(\ref{cureqs}), we first transform our variables into Fourier space as follows $a_{kL} = 1/\sqrt{\Omega}\sum_{l}e^{-i(k-\phi/2)l}a_{lL}$ and $a_{kR} = 1/\sqrt{\Omega}\sum_{l}e^{-i(k+\phi/2)l}a_{lR}$. Inserting these expressions into Eq.~(\ref{cureqs}) and expressing $a_{k\mu}$ in terms of $\beta_{k}$, we obtain:
\begin{equation}\label{rungcur}
j^{\perp}_{l} = -\frac{K}{\Omega}\sin{2\gamma}\sin{2k_{0}l}\cos{\theta_{k_{0}}}
\end{equation}
The ladder currents are similarly found to be:
\begin{eqnarray}\label{ladcur}
j^{\parallel}_{lL} = \frac{J}{\Omega}(2\sin^{2}{\frac{\theta_{k_{0}}}{2}}\sin{(k_{0} + \frac{\phi}{2})}\cos^{2}{\gamma} + \\\nonumber 2\cos^{2}{\frac{\theta_{k_{0}}}{2}}\sin{(-k_{0} + \frac{\phi}{2})}\sin^{2}{\gamma} +\\\nonumber  i\frac{\sin{\theta_{k_{0}}}\sin{\gamma}}{4}\Big(e^{-2ik_{0}l}(e^{-i(k_{0} + \frac{\phi}{2})} - e^{-i(k_{0} - \frac{\phi}{2})}) -\\\nonumber \text{c.c}\Big) \\\nonumber
j^{\parallel}_{lR} = \frac{J}{\Omega}(-2\cos^{2}{\frac{\theta_{k_{0}}}{2}}\sin{(-k_{0} + \frac{\phi}{2})}\cos^{2}{\gamma} - \\\nonumber 2\sin^{2}{\frac{\theta_{k_{0}}}{2}}\sin{(k_{0} + \frac{\phi}{2})}\sin^{2}{\gamma} +\\\nonumber  i\frac{\sin{\theta_{k_{0}}}\sin{\gamma}}{4}\Big(e^{-2ik_{0}l}(e^{-i(k_{0} - \frac{\phi}{2})} - e^{-i(k_{0} + \frac{\phi}{2})}) -\\\nonumber \text{c.c}\Big)
\end{eqnarray}

The reader may readily check that using the fact that there is a symmetry from $k_{0} \rightarrow -k_{0}$, and that $\cos^{2}{\frac{\theta_{-k_{0}}}{2}} = \sin^{2}{\frac{\theta_{k_{0}}}{2}}$, the net ladder current $j^{\parallel}_{lL} + j^{\parallel}_{lR}$ vanishes on every site $l$, and the net chiral current is $j_{c} = 
\sum_{l}(j^{\parallel}_{lL} - j^{\parallel}_{lR}) = 4J\sin^{2}{\frac{\theta_{k_{0}}}{2}}\sin{(k_{0}+\phi/2)}$.


\end{appendix}

\bibliography{dipolarbib}

\end{document}